\documentclass[%
 aps,
 prl,%
 amsmath,amssymb,floatfix,
 reprint,superscriptaddress,%
]{revtex4-1}

\usepackage[version=3]{mhchem} %
\usepackage[T1]{fontenc}       %
\usepackage{color}
\usepackage[normalem]{ulem}
\usepackage{graphicx}

\begin{document}

\author{Chungwen Liang}
\email{chungwen.liang@gmail.com}
\affiliation{Laboratory of Computational Science and Modeling, Institute of Materials, {\'E}cole Polytechnique F{\'e}d{\'e}rale de Lausanne, 1015 Lausanne, Switzerland}
\affiliation{Laboratory for fundamental BioPhotonics, Institutes of Bioengineering and Materials Science and Engineering, School of Engineering, and Lausanne Centre for Ultrafast Science,  {\'E}cole Polytechnique F{\'e}d{\'e}rale de Lausanne, CH-1015 Lausanne, Switzerland}

\author{Gabriele Tocci}
\affiliation{Laboratory of Computational Science and Modeling, Institute of Materials, {\'E}cole Polytechnique F{\'e}d{\'e}rale de Lausanne, 1015 Lausanne, Switzerland}
\affiliation{Laboratory for fundamental BioPhotonics, Institutes of Bioengineering and Materials Science and Engineering, School of Engineering, and Lausanne Centre for Ultrafast Science,  {\'E}cole Polytechnique F{\'e}d{\'e}rale de Lausanne, CH-1015 Lausanne, Switzerland}

\author{David M. Wilkins}
\affiliation{Laboratory of Computational Science and Modeling, Institute of Materials, {\'E}cole Polytechnique F{\'e}d{\'e}rale de Lausanne, 1015 Lausanne, Switzerland}
\affiliation{Laboratory for fundamental BioPhotonics, Institutes of Bioengineering and Materials Science and Engineering, School of Engineering, and Lausanne Centre for Ultrafast Science,  {\'E}cole Polytechnique F{\'e}d{\'e}rale de Lausanne, CH-1015 Lausanne, Switzerland}

\author{Andrea Grisafi}
\affiliation{Laboratory of Computational Science and Modeling, Institute of Materials, {\'E}cole Polytechnique F{\'e}d{\'e}rale de Lausanne, 1015 Lausanne, Switzerland}

\author{Sylvie Roke}
\affiliation{Laboratory for fundamental BioPhotonics, Institutes of Bioengineering and Materials Science and Engineering, School of Engineering, and Lausanne Centre for Ultrafast Science,  {\'E}cole Polytechnique F{\'e}d{\'e}rale de Lausanne, CH-1015 Lausanne, Switzerland}

\author{Michele Ceriotti}
\affiliation{Laboratory of Computational Science and Modeling, Institute of Materials, {\'E}cole Polytechnique F{\'e}d{\'e}rale de Lausanne, 1015 Lausanne, Switzerland}

\title{Solvent Fluctuations and Nuclear Quantum Effects \\
Modulate the Molecular Hyperpolarizability of Water}


\begin{abstract}
Second-Harmonic Scatteringh (SHS) experiments
provide a unique approach to probe
non-centrosymmetric environments in aqueous
media, from bulk solutions to interfaces, living cells and tissue. A central assumption
made in analyzing SHS experiments is that 
the each molecule scatters light according to a constant molecular hyperpolarizability tensor
$\boldsymbol{\beta}^{(2)}$.
Here, we investigate the dependence of the molecular
hyperpolarizability of water on its environment and
internal geometric distortions, in order to test
the hypothesis of constant $\boldsymbol{\beta}^{(2)}$.
We use quantum chemistry calculations of
the hyperpolarizability of a molecule
embedded in point-charge environments 
obtained from simulations of bulk water.
We demonstrate that both the heterogeneity
of the solvent configurations and the
quantum mechanical fluctuations of the
molecular geometry introduce large
variations in the non-linear optical
response of water. This finding has the
potential to change the way SHS experiments
are interpreted: in particular, isotopic
differences between \ce{H2O} and \ce{D2O} 
could explain recent SHG scattering observations.
Finally, we show that a 
simple machine-learning framework can predict
accurately the fluctuations of the molecular 
hyperpolarizability.
This model accounts for the microscopic inhomogeneity of the solvent
and represents a first step
towards quantitative modelling of SHS 
experiments.

\end{abstract}
\maketitle

Nonlinear optical (NLO) processes are of great interest in physics, 
chemistry, biology and materials science, as they provide a means of
probing the structure and behavior of liquids, 
nanostructures and interfaces \cite{Boyd,Shen}.
Second harmonic generation (SHG) is a NLO process 
in which two photons with frequency $\omega$ are instantaneously combined to generate new photons with frequency 2$\omega$ after interacting with a material. As a second-order NLO process, SHG is only allowed in non-centrosymmetric environments. SHG spectroscopy experiments in molecular systems can be carried out in three different geometries: reflection, transmission, and scattering (SHS) \cite{Shen1989,Eisenthal2005,Roke2012}. The properties of planar interfaces are often probed by SHG spectroscopy in the reflection mode, while the properties of 
spherical interfaces and bulk materials are often probed by SHG 
spectroscopy in the scattering mode \cite{Terhune1965,Scheu2014}. 
The structural information of molecular systems, such as molecular 
adsorption and orientation on metal surfaces \cite{Shen1989,Geiger2009}, 
polarity of liquid interfaces \cite{Wang1997}, nanoparticles in solutions 
\cite{Liu1999,Butet2010} and bulk molecular liquids 
\cite{Shelton2012}, has been intensively studied by SHG 
spectroscopy. 

Theoretical frameworks for estimation of the SHG response in the 
reflection and scattering modes have long been known
\cite{Bersohn1966,Sokhan1997}, and are necessary to interpret
experimental results. 
The SHS response of a molecular system simultaneously carries information 
on the structural correlations and the nonlinear optical response of 
each molecule, and modelling is required to disentangle these 
contributions to the experimental measurements\cite{Galli2015}. 
However, it is challenging to do so without introducing harsh approximations. For instance, to extract information on orientational correlations at
interfaces or in the bulk phase,
it is common to assume that scattering from molecules
in solution is incoherent \cite{Terhune1965,Kauranen1996,Wang1997}.
However, recent experiments and simulations have found evidence of a
significant coherent contribution to the scattering, particularly
in the case of hydrogen-bonded solvents.
\cite{Shelton2014,Shelton2015,Chen2016,Tocci2016}
Another critical assumption that is often made is that
the hyperpolarizability tensor $\boldsymbol{\beta}^{(2)}$ of the molecules
is constant, independent of the environment and the molecular
geometries. Most experimental analyses rely on electronic structure
calculations to obtain an estimate of $\boldsymbol{\beta}^{(2)}$.
Early computational studies focused on calculating this
microscopic quantity for gas-phase molecules \cite{Kurtz1990,Maroulis1991,Bartlett1993},
while more recently the role of solvation has also been considered \cite{Mikkelsen1994,Bella1994,Kusalik2001}.

Due to its ubiquitous presence in chemical
and biological systems, water has been 
given special attention in experimental and theoretical SHS studies.
As a consequence of the strong electrostatic interactions
in the liquid, the electronic structure of
water and therefore its molecular hyperpolarizability change dramatically on going from the gas phase to the liquid phase \cite{Levine1976,Ward1979}.
To determine these changes quantitatively, several quantum chemistry calculations have been performed based on simple point charge environments \cite{Kusalik2001}, dielectric continuum theories \cite{Kongsted2003}, solvation models \cite{Sylvester2004}, and mixed quantum/classical (QM/MM) approaches \citep{Kongsted2003,Snijders2003} to incorporate the environmental effect.
Even though the value of the hyperpolarizability is very sensitive to the level of theory, functional and basis set, 
all of these studies report a sign change of the elements of $\boldsymbol{\beta}^{(2)}$ upon changing from a gas phase environment to the liquid phase. Despite the fact that a strong dependence on the molecular configurations used in the calculations has been reported \cite{Garbuio2009}, most theoretical studies have assumed that the water hyperpolarizability tensor elements are constant \cite{Kusalik2001,Kongsted2003,Sylvester2004,Snijders2003,Shiratori2013}, and thus independent of the inhomogeneous liquid environment or the internal geometry of the water molecule. 
It should, however, be noted that in the most commonly adopted
description the SHS process is assumed to take place instantaneously, so that each water molecule should respond according to its environment. Only by simultaneously taking into account the structural correlations between molecules \cite{Shiratori2013,Tocci2016}
and the variation of their
second-harmonic response would it be possible to reach 
an approximate
quantitative description of SHS experiments.

In this paper we investigate the hyperpolarizability of water molecules in the liquid phase,
and demonstrate that the inhomogeneous electrostatic environment
has a dramatic impact on the elements of $\boldsymbol{\beta}^{(2)}$. We also consider the role played by thermal and quantum
fluctuations of the internal coordinates of each molecule,
finding evidence for a significant isotope effect
between \ce{H2O} and \ce{D2O}.
Finally, we establish a theoretical framework that allows us to
combine an accurate quantum mechanical evaluation of
the second-order response with a machine-learning model
that can accurately predict the behavior of molecules in
large-scale molecular dynamics simulations.
We envision that this framework will facilitate the calculation of the full SHS intensity from atomistic simulations, which we leave for future work.

In order to investigate the role of solvent fluctuations
in determining the hyperpolarizability of a water molecule
in the liquid phase we
use an embedding approach inspired by QM/MM methods,
where the hyperpolarizability of a central water
molecule is treated quantum mechanically, whereas the
surrounding molecules are treated classically.
We first perform extensive, long-time and large-scale molecular dynamics  (MD) simulations of
bulk liquid water using
fixed point charge models \cite{TIP4P-2005,q-tip4p} (see Supporting Information for the simulation details). From the results of these 
simulations we extract random configurations of water environments, by taking 
molecules within 1.5 nm of a 
central water molecule. We show in the SI that this cutoff is
sufficient to provide a representative sampling of the electrostatic
environment in bulk water.
We perform quantum chemistry calculations of the 
hyperpolarizability tensor of the central molecule,
with the surrounding molecules modelled as
point charges consistent with the empirical force-field.
Since our objective here is to
assess the importance of 
fluctuations on the molecular
hyperpolarizability of water, and 
to develop a computational framework
that is compatible with the large-scale
simulations needed to model SHS 
experiments, we limit our 
discussion to this simple monomer embedding. All hyperpolarizability calculations were performed 
at the CCSD/d-aug-cc-pvtz level using the Dalton 2015 package \cite{Dalton}. 
The hyperpolarizability tensor element $\beta_{ijk}$ is given 
by the numerical derivative of the energy $U$ with respect 
to the external electric fields $E_{i}$, $E_{j}$, and $E_{k}$: 
\begin{equation}
\beta_{ijk}=\frac{\partial^{3} U}{\partial E_{i} \partial E_{j} \partial E_{k}}.
\end{equation}
We calculate the static hyperpolarizability tensor, which is an approximation to the full frequency dependent tensor probed in SHS experiments.
For aqueous systems at the frequencies typically used in elastic second-harmonic scattering experiments this approximation
can be expected to entail an error smaller than 10\%~\cite{Sylvester2004} -- which should not 
affect the qualitative scope of our discussion of
the assessment and the machine-learning of the 
local fluctuations of $\boldsymbol{\beta}^{(2)}$.

\begin{figure}[!t]
\includegraphics[width=7 cm]{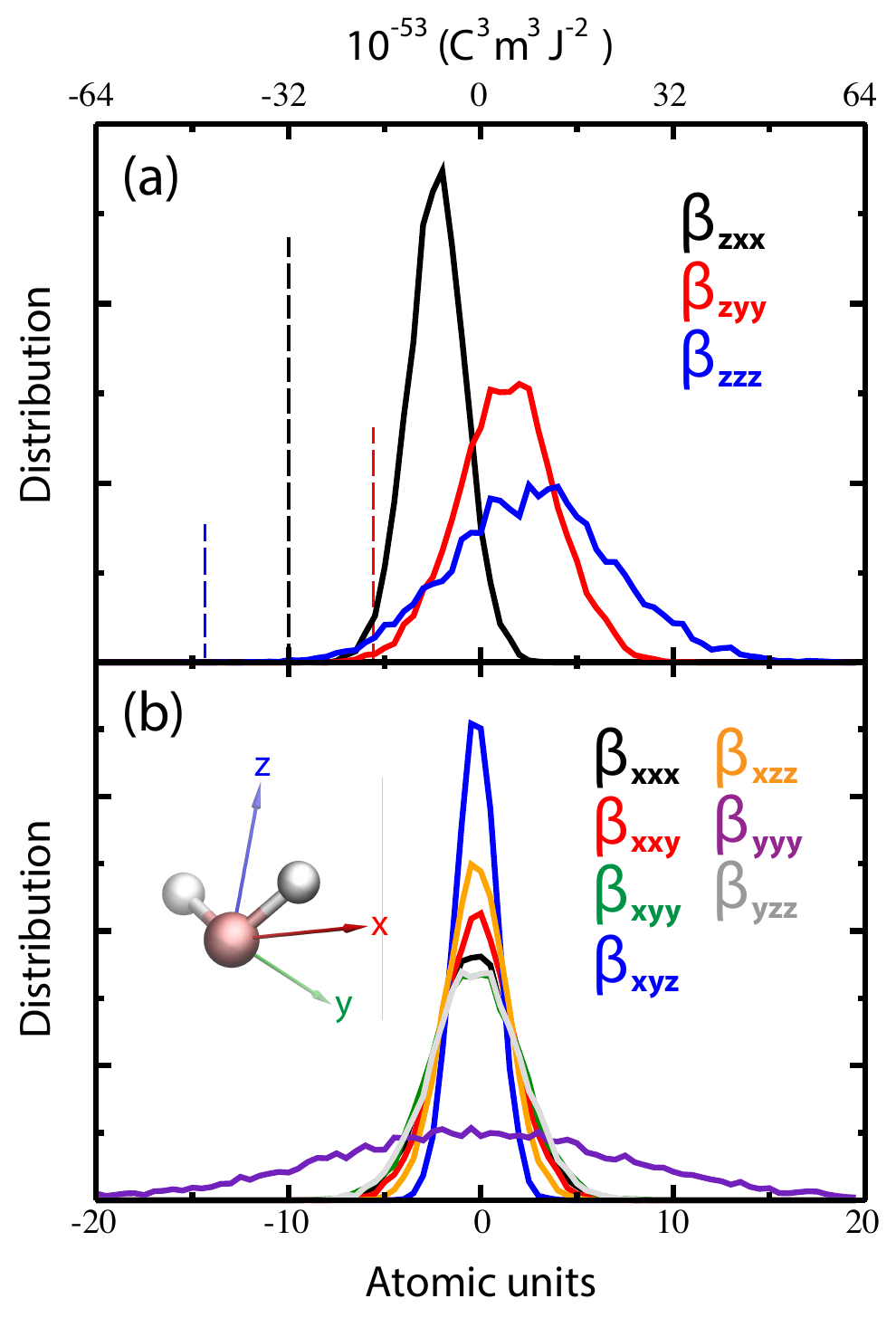}
\caption{\label{fig:tensor_distributions}%
(a) The distributions of three tensor elements $\beta_{zxx}$, $\beta_{zyy}$ and $\beta_{zzz}$. For comparison, the constant gas phase values are shown as dashed lines. (b) The distributions of the remaining seven tensor elements. Inset: the orientation of the central water molecule.} 
\end{figure}

The distributions of  $\beta_{zxx}$, $\beta_{zyy}$ and $\beta_{zzz}$ are shown in Figure \ref{fig:tensor_distributions}(a), and compared
to the values for a (rigid) gas-phase molecule. 
It can be seen that
these elements have a wide distribution and 
are shifted towards positive values compared
to the gas-phase.
Both effects are most pronounced for
$\beta_{zzz}$.
Furthermore, it should be noted that most 
previous studies have focused on calculations of these three elements, 
which are the only independent, non-zero values considering the $C_{2v}$ symmetry of a water molecule~\cite{Giordmaine1965,Bersohn1966}.
Fluctuations in the liquid phase break 
this symmetry, so that instantaneously $\boldsymbol{\beta}^{(2)}$ has 10 independent non-zero elements. 
Figure \ref{fig:tensor_distributions}(b) shows the distributions
of the tensor elements that would be zero under $C_{2v}$ symmetry. While the average of these elements vanishes, their spread is comparable to that 
of $\beta_{zxx}$ -- and much larger in the case of $\beta_{yyy}$ -- suggesting that these elements may contribute significantly
to the total SHS response of aqueous systems.
This figure clearly shows that neglecting
environmental fluctuations and treating the
hyperpolarizability as a constant
constitutes a severe approximation, and may have an effect on the interpretation of experiments.

\begin{figure}[hbt]
\includegraphics[width=8.2 cm]{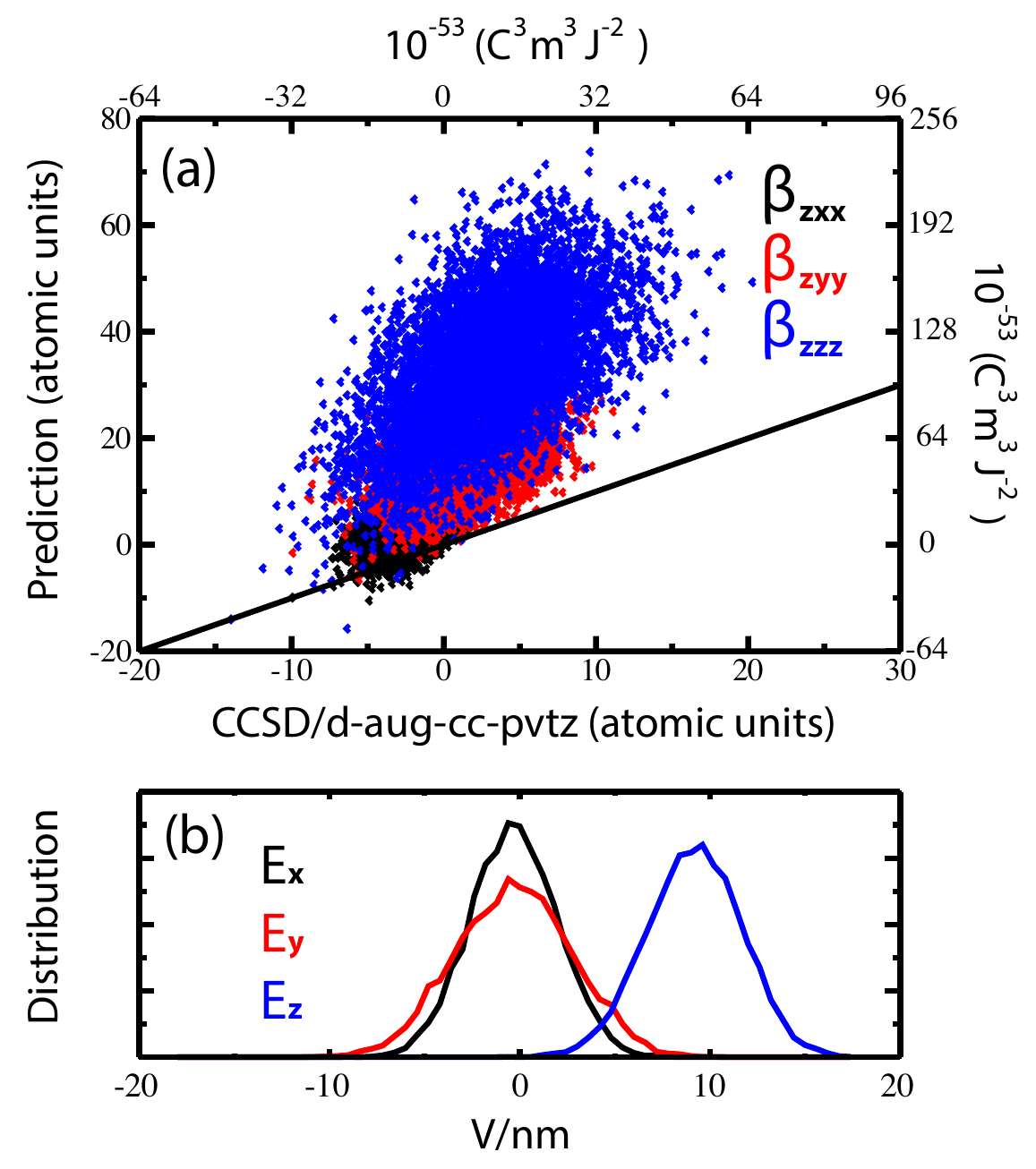}
\caption{ \label{fig:simple_model}
(a) Correlation plot between the molecular hyperpolarizability elements calculated using quantum chemistry and those estimated using Eqn.~\eqref{eq:taylor}.
(b) Distribution of electric fields along the $x$, $y$ and $z$ directions in the molecular frame.} 
\end{figure}

Let us now consider the physical
origin of these fluctuations, and of 
the positive shift of $\beta_{zxx}$, $\beta_{zyy}$ and $\beta_{zzz}$. If one assumes that the overall hyperpolarizability can be described by a Taylor expansion of higher-order polarizabilities which couple with the local electric fields, a tensor element $\beta_{ijk}^{\text{liquid}}$ in the liquid phase can be written as:
\begin{equation}
\beta_{ijk}^{\text{liquid}} = \beta_{ijk}^{\text{gas}} + \sum_{l=x,y,z} \gamma_{ijkl}^{\text{gas}} E_{l} 
\label{eq:taylor}
\end{equation}
where $\gamma_{ijkl}^{\text{gas}}$ is the tensor element of the water third-order polarizability ($\boldsymbol{\gamma}^{(3)}$) in the gas phase. $E_{l}$ is the electric field along the $x$, $y$ or $z$ direction evaluated at the position of the
O atom of the central water molecule. The contribution of the higher order hyperpolarizabilities is assumed to be negligible.
To rule
out contributions from the distortions
of each monomer, we will
consider snapshots from our simulation
of rigid TIP4P/2005 water.
$\gamma_{ijkl}^{\text{gas}}$ is calculated based on the geometry of the TIP4P/2005 water model (shown in Table S1). The correlation plots of the values of $\beta_{zxx}$, $\beta_{zyy}$ and $\beta_{zzz}$ computed based on the embedded monomer model, and those estimated  from Eqn.~\eqref{eq:taylor},  %
are shown in Figure \ref{fig:simple_model} (a), together
with the distributions of the electric field components, shown in Figure \ref{fig:simple_model} (b). 
We show in Table S1 that the tensor elements $\gamma_{zxxz}^{\text{gas}}$, $\gamma_{zyyz}^{\text{gas}}$, and $\gamma_{zzzz}^{\text{gas}}$ are large positive numbers, while the other components are near-zero. Hence, 
$E_{x}$ and $E_{y}$ contribute negligibly to the shift, while 
the electric field along the water dipole direction $E_{z}$ (which 
generally takes positive values) is
predicted to induce a positive shift
on the values of  $\beta_{zxx}$, $\beta_{zyy}$, and $\beta_{zzz}$. 
Similar considerations also apply to the
other elements of $\boldsymbol{\beta}^{(2)}$.
For instance the large fluctuations in $\beta_{yyy}$
result from the large value of $\gamma^\text{gas}_{yyyy}$
and the large spread in $E_y$. 
While the values of the gas-phase $\boldsymbol{\gamma}^{(3)}$ and of 
the local electric field explain the
qualitative shift of $\boldsymbol{\beta}^{(2)}$ upon condensation,
it is clear that the simple model
in Eqn.~\eqref{eq:taylor} is insufficient to quantitatively predict 
the molecular response of water.

\begin{figure}[hbt]
\includegraphics[width=1.0\columnwidth]{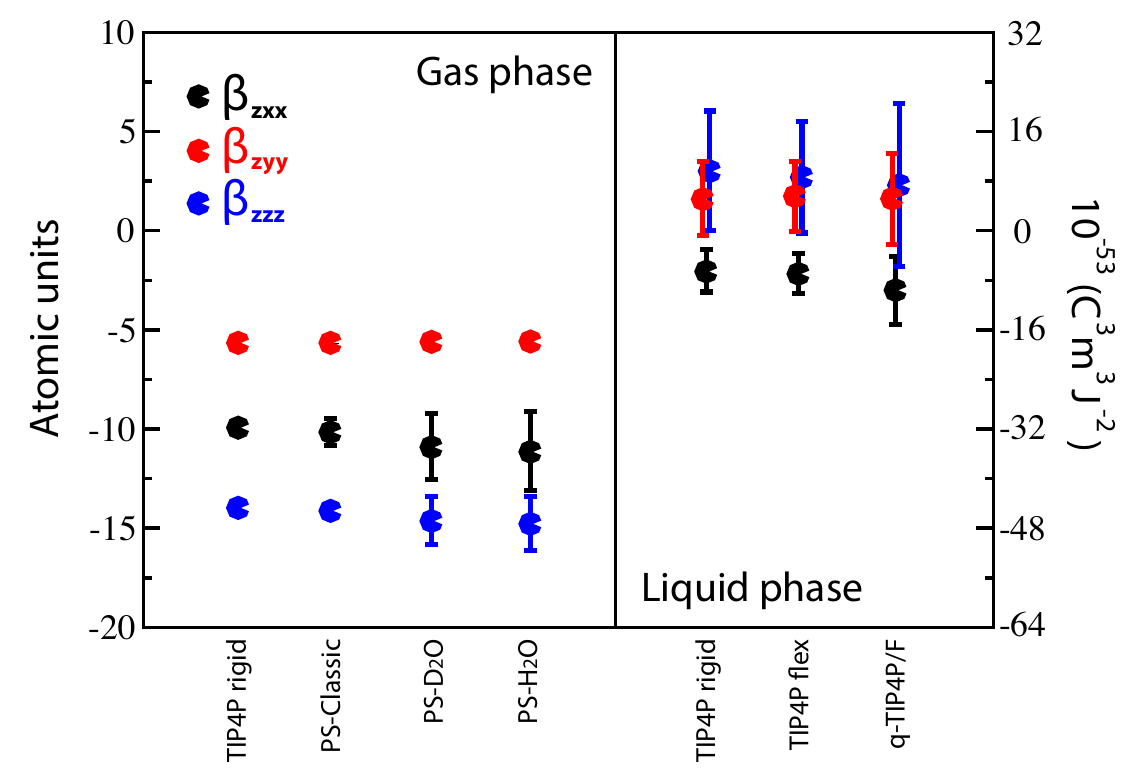}
\caption{\label{fig:flexible} 
The mean and standard deviation of three tensor elements $\beta_{zxx}$, $\beta_{zyy}$ and $\beta_{zzz}$ calculated in the gas phase with the rigid TIP4P/2005 model for rigid, classical \ce{H2O} and the Partridge-Schwenke monomer potential for classical \ce{H2O} and quantum \ce{D2O} and \ce{H2O}, and in the liquid phase with the TIP4P/2005 and TIP4P/2005-flexible models for classical \ce{H2O} and with the q-TIP4P/F potential for quantum \ce{H2O}.
The bars represent the intrinsic variation of the molecular response due to differences in environments and internal distortions. Statistical errors are about 1\% of the standard deviation.
}
\end{figure}

Before discussing how a more accurate model 
can be constructed, let us consider how
molecular distortions and
nuclear quantization affect $\boldsymbol{\beta}^{(2)}$.
To this aim, we 
carry out calculations in the liquid phase
with two flexible water models: classical MD simulations with TIP4P/2005-flexible \cite{tip4p2005f} and path integral MD (PIMD) simulations with q-TIP4P/F \cite{q-tip4p}
-- which have parametrizations essentially equivalent to the rigid TIP4P/2005, 
fitted to reproduce the structural and vibrational properties of water with classical and quantum statistics. 
In order to evaluate high-accuracy reference
values for the gas phase, we also perform classical MD and high-order PIMD~\cite{kapi+16jcp2} 
simulations using the Partridge-Schwenke monomer potential~\cite{part-schw97jcp}. 
The details of the classical MD and the PIMD simulations are described in the SI. 
Following the same procedure as before, we extracted 10,000 water clusters from the trajectories.
The mean and the standard deviation of the
three tensor elements $\beta_{zxx}$,  $\beta_{zyy}$ and $\beta_{zzz}$ calculated from the TIP4P/2005-flexible and q-TIP4P/F models are shown in Figure \ref{fig:flexible}.
For classical water in both the liquid and in the gas phase,
thermal fluctuations of the
molecular geometry at 300 K lead 
to negligible changes in the 
distribution of the elements of
$\boldsymbol{\beta}^{(2)}$.
However,
when nuclear quantum effects are 
introduced, the distributions of the 
$\boldsymbol{\beta}^{(2)}$ tensor 
elements are considerably broadened, with a standard deviation for the $\beta_{zzz}$
component of the q-TIP4P/F
model that is approximately 30\% larger than its classical counterparts.
This observation is consistent
with the large changes that are seen
in the electronic properties of water
when nuclear quantum effects are properly accounted for, e.g. the 
band gap~\cite{gibe+14jpcb,chen+16prl}
or the H-NMR chemical shifts~\cite{ceri+13pnas}, which are
connected to the increased delocalization of the proton along the
H-bond.

Significant fluctuations of 
$\boldsymbol{\beta}^{(2)}$ are also seen for the gas-phase
simulations, 
stressing that internal molecular fluctuations
modulate the molecular response, on an 
ultra-fast timescale. From our calculations, we can
extract the mean value
of $\beta_{\parallel} = \frac{3}{5}\left(\beta_{zxx}+\beta_{zyy}+\beta_{zzz}\right)$, which is a measurable quantity in electric field induced second harmonic generation (EFISHG) experiments \cite{Levine1976,Ward1979}. The
value we obtain -- $\langle\beta_{\parallel}\rangle=-18.93 (-18.69)$ a.u., for \ce{H2O}(\ce{D2O}) -- agree very well
with the experimental results reported in 
Ref.~\cite{kaat+98jcp} of $-19.2\pm 0.9$($-17.8\pm1.2$) a.u.
and in Ref.~\cite{Ward1979} of
$-22.0\pm0.9$ a. u. for \ce{H2O}. 
This shows that quantum fluctuations have a
pronounced effect on the molecular
hyperpolarizability in both the
gas and liquid phases.
Results for liquid water --
$\beta_{\parallel}= 1.53 (0.54)$ a.u. for 
classical(quantum) \ce{H2O} -- show
significant deviation from the experimental value of 3.19 a.u.,\cite{Levine1976}
but are
much closer than the commonly adopted
values from fixed environments -- which can be
as high as 16.3 a.u. \cite{Kusalik2001}.

Having assessed the role of
quantum fluctuations and 
that of the inhomogeneous
environment in determining
the values of  $\boldsymbol{\beta}^{(2)}$
we now design a
machine learning model for
the prediction of
$\boldsymbol{\beta}^{(2)}$
in the liquid phase.
This model incorporates the dependence of
$\boldsymbol{\beta}^{(2)}$ on the
inhomogeneous environment and
on quantum fluctuations, and 
is an essential requirement for the development of
a framework to compute
the SHS response of liquid water
from MD simulations without performing computationally demanding quantum chemistry calculations.

\begin{figure}[hbt]
\includegraphics[width=7.3 cm]{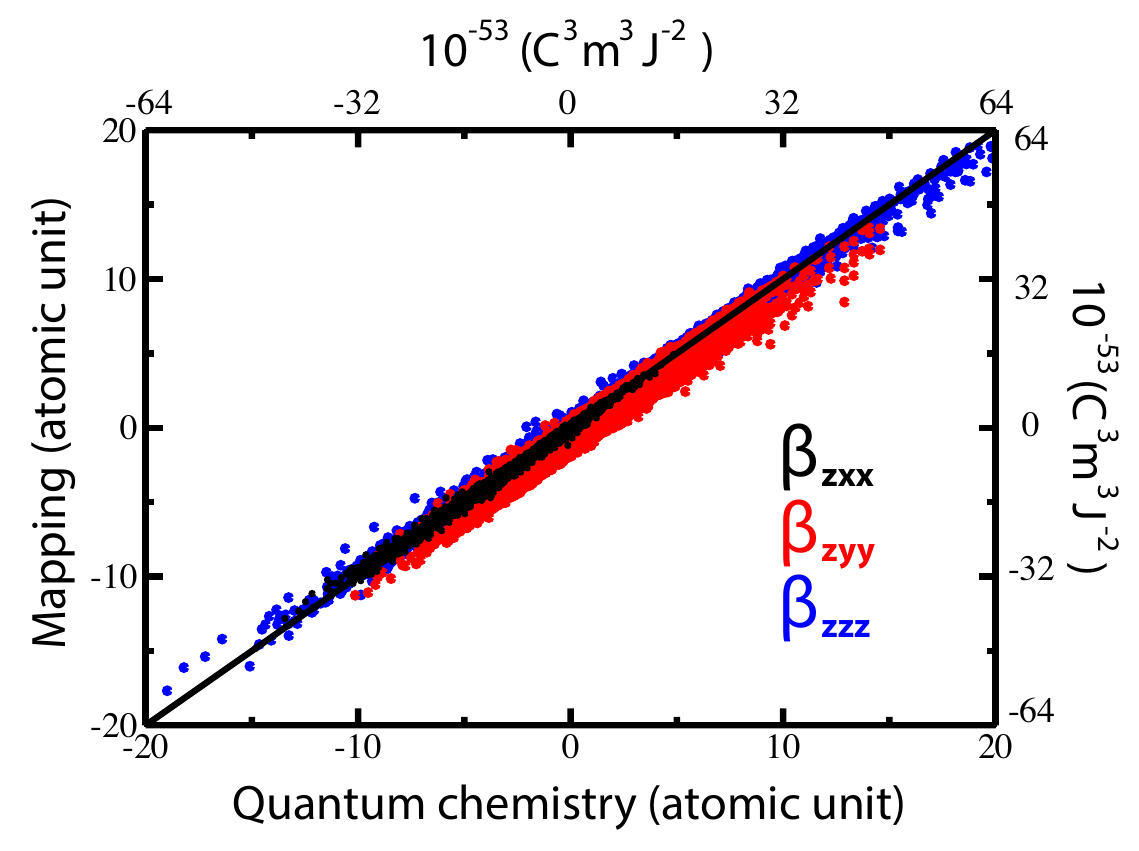}
\caption{The correlation plots of $\beta_{zxx}$,  $\beta_{zyy}$ and $\beta_{zzz}$ between quantum chemistry calculations and the
machine-learning mapping procedure. } \label{fig:krr}
\end{figure}

The construction of the machine learning model
and the selection of hyperparameters
is described in detail in the 
Supporting Information:
we define a grid of points
surrounding a central water molecule.
Inspired by the observations discussed
above, we describe each environment by a 
vector $\mathbf{u}$ that contains both 
the electric field generated by
all water molecules in the environment,
and a smooth Gaussian representation of the
oxygen and hydrogen atom densities \cite{bart+13prb}, which
accounts for the
dependence of the hyperpolarizability on
short-range interactions and
molecular distortions. We adopt a kernel ridge regression model to learn the hyperpolarizabilities computed from quantum chemistry \cite{scho+98nc}:

\begin{equation}
\beta_{ijk}(\mathbf{u}) = \bar{b}^{(ijk)} + \sum_l c^{(ijk)}_l K\left(\mathbf{u},\mathbf{u}_l\right) 
\label{eq:krr}
\end{equation}
where we use a Gaussian kernel $K\left(\mathbf{u},\mathbf{u}'\right)=e^{-\left|\mathbf{u}-\mathbf{u}'\right|^2/\sigma^2}$
and optimize the weights $c^{(ijk)}_l$ by 
minimizing the prediction error for a training set. 
Once the weights have been determined, one can 
easily predict the components of $\boldsymbol{\beta}^{(2)}$ 
using Eqn.~\eqref{eq:krr}.
As shown in Fig.~\ref{fig:krr}, 
the model can predict
the different components of the
hyperpolarizability tensor for a test set, with a RMS error of 6\%.

In summary, we have demonstrated that
the hyperpolarizability of liquid water
fluctuates significantly
due to the inhomogeneities of the local
molecular environment and to nuclear 
quantum effects.
In doing so, we build on previous work that shows the dependence of water's hyperpolarizability
on its environment \cite{Kusalik2001,Kongsted2003,Sylvester2004}, by 
explicitly considering the
H-bonding fluctuations of this environment.
We see that the assumption of a
constant molecular $\boldsymbol{\beta}^{(2)}$,
commonly adopted in interpreting SHS
experiments, needs to be revised.
Fluctuations in the hyperpolarizability enter naturally into the analysis of second harmonic experiments, because the expression for the second harmonic intensity contains terms that depend on the square of elements of $\boldsymbol{\beta}^{(2)}$ \cite{Shen1989,Roke2012}.
By providing a quantitative estimate of these fluctuations our work 
may aid the interpretation of SHS experiments.
Although our results concern bulk water, fluctuations in $\boldsymbol{\beta}^{(2)}$ are present
also in interfacial and inhomogeneous systems, which
are more relevant to second harmonic
experiments.
Including  the effects of environmental, geometric, and
nuclear quantum fluctuations gives a molecular
tensor that agrees much better with the results
of the EFISHG experiments than previous work --
reaching quantitative agreement in the gas phase.
The isotopic dependence of the molecular response is 
particularly intriguing, as this could
contribute to the explanation of recent 
experimental findings~\cite{Chen2016} showing
that the SHS signal from dilute ionic 
solutions is largely non-ion specific, 
but varies dramatically with the isotopic 
composition of the solvent (H$_{2}$O vs. 
D$_{2}$O).
To achieve quantitative modelling
of SHS and answer these questions,
it is desirable to calculate 
the SHS response of the system directly 
from MD trajectories, going beyond the
approximation of a constant molecular 
response.  We take steps towards this goal
by introducing a machine-learning
framework that can predict the fluctuations
in molecular response without needing to
resort to expensive quantum chemistry 
calculations. Although we have applied it only
to bulk water, this framework can be extended to
general systems,
making it a powerful tool for the study of interfaces.
We show that  the full 
response tensor
can be approximated by an
embedded-monomer model, although 
many-body effects are important and could
be included as a further refinement. While it is possible 
to model nanosecond experiments using 
a constant mean-field value for 
$\boldsymbol{\beta}^{(2)}$~\cite{Tocci2016},
the realization that on a molecular level the 
hyperpolarizability reflects the interplay between 
quantum mechanical and electrostatic fluctuations
opens up the possibility of using ultrafast SHS
experiments to probe these effects. 
Future work will involve the quantitative simulation of 
SHG measurements, which will greatly increase the 
interpretative power of non-linear optical
experiments of complex aqueous systems,
including the study of ion absorption on the water surface \cite{Otten2012},
the assessment of 
molecular orientation 
at the air/water interface \cite{Kundu2013}, 
and the structure of surfactant molecules interacting with 
nanoparticles \cite{You2012}.

\section*{Acknowledgement}
C.L., G.T., and S.R. are grateful for support from
the Julia Jacobi Foundation and the European Research Council (Project Number 616305).
D.M.W. and M.C. acknowledge the Swiss National Science Foundation 
(Project ID 200021\_163210).
This work was supported by a grant from the Swiss National Supercomputing
Centre (CSCS) under project ID s619.

\end{document}